# DISTRIBUTION AND CONTENT OF DUST IN OVERLAPPING GALAXY SYSTEMS

R.E. WHITE III AND W.C. KEEL
*University of Alabama*
*Department of Physics & Astronomy*
*Tuscaloosa, AL 35487-0324 USA*

AND

C.J. CONSELICE
*University of Chicago*
*Department of Astronomy & Astrophysics*
*5640 S. Ellis Avenue*
*Chicago, IL 60637 USA*

## 1. Introduction

We present a progress report on our program to determine the opacity of spiral disks *directly*, rather than statistically, by imaging foreground spirals partially projected against background galaxies. The non-overlapping regions of partially overlapping galaxies can be used to reconstruct, using purely differential photometry, how much light from the background galaxy is lost in passing through the foreground galaxy in the overlap region.

There are several benefits to the direct, differential photometric approach we have adopted, including: 1) it is not subject to the selection effects influencing statistical studies; 2) there is no selection against high opacity regions; 3) the imaging technique involves only differential photometry; 4) large, contiguous areas can be analyzed, allowing average values of the opacity to be estimated; 5) there is no need to correct for the internal extinction of the background galaxy or the Milky Way; 6) scattering corrections are also differential, which can keep them slight.

This technique also has some disadvantages relative to others: 1) there are rather few tractable objects nearby enough for spatially well-resolved analysis; and 2) the success of the technique hinges on the degree of symmetry in both the foreground and background galaxies

*Figure 1.*  The ideal overlapping galaxy pair for the direct opacity measures.

Figure 1 depicts the ideal case for constructing maps of opacity using differential photometry: a foreground disk (spiral) galaxy is half-projected against half of a similarly-sized background elliptical galaxy. For the sake of illustration, the (unobscured) surface brightness of each galaxy is taken to be constant, with $F$ and $B$ being the actual surface brightness values of the foreground disk and background elliptical in the overlap region, and $\tau$ is the optical depth in the disk. The observed surface brightness in the overlap region is then $\langle F + Be^{-\tau}\rangle$, where brackets are used to emphasize that this whole quantity is the observable in the overlap region and cannot be directly decomposed into its constituent components. We use symmetric counterparts from the non-overlapping regions of the two galaxies to *estimate* $F$ and $B$ and denote the estimates as $F'$ and $B'$. We can then construct an estimate of the optical depth, denoted $\tau'$:

$$e^{-\tau'} = \frac{\langle F + Be^{-\tau}\rangle - F'}{B'}. \qquad (1)$$

Here the estimate of the foreground spiral's surface brightness, $F'$, is first subtracted from the surface brightness of the overlap region, $\langle F + Be^{-\tau}\rangle$; this result is then divided by the estimate of the background elliptical's surface brightness, $E'$. This creates a map of $e^{-\tau'}$ in the overlap region.

## 2. Sample Selection and Results

Our observing sample is largely drawn from numerical searches for overlapping neighbors in the *RC3, RSA, ESO-Uppsala, UGC, RNGC, NGC2000, MCG*, Karachentsev, and Zhenlong et al (1989) catalogues. We also inspected individual catalog entries in the *UGC*, *ESO-LV*, and *NGC* listings which were typed as inherently multiple systems. In addition, we visually inspected all pairs in the Arp-Madore catalog, the Arp *Atlas of Peculiar*

*Figure 2.* Six galaxy pairs from which extinctions have been deduced.

*Galaxies*, and the Reduzzi & Rampazzo (1995) catalog of southern pairs. Figure 2 shows six of the dozen or so galaxy pairs that we have successfully analyzed after imaging them in the $B$ and $I$ bands.

## 3. Results

Table 1 lists the extinctions determined in arm and interarm regions in various backlit galaxies. For each measurement, we list its radial distance from the center of the foreground galaxy in units of $R_{25}^B$. Typical errors in the extinctions are $\sim 0.15$ mag. Figure 3 plots the contents of Table 1.

## 4. Conclusions

We generally find that opacity is concentrated in spiral arms, while interarm regions are nearly transparent, particularly beyond $\sim 0.5 R_{25}^B$. This confirms and extends our previous results (White & Keel 1992, Keel & White 1995). Since most disk light comes from spiral arms, the dust opacity is correlated with the emission. This correlation may explain why the surface brightness of spirals is found to be independent of inclination in statistical tests (Valentijn 1991), erroneously implying that spiral disks are generally opaque.

This project was supported by the *NSF* and the State of Alabama through the *EPSCoR* program, by the *NSF Research Experience for Undergraduates* program, and by Lowell Observatory and NOAO.

TABLE 1. Face-on-corrected Magnitudes of Extinction

| Galaxy Pair | arm | | | interarm | | |
|---|---|---|---|---|---|---|
| | $R/R_{25}^B$ | $A_B$ | $A_I$ | $R/R_{25}^B$ | $A_B$ | $A_I$ |
| AM 0500-620 | 0.6 | > 2.3 | 1.64 | 0.5 | 0.1-0.47 | 0.0-0.55 |
| AM 1311-455 | 1.18 | 0.73 | 0.24 | 0.95 | 0 | 0 |
| AM 1316-241 | 0.75 | 0.38 | 0.16 | 0.4-0.75 | 0.08 | 0.05 |
| ESO 0320-51 | 0.65 | 0.27 | 0.17 | 0.50 | 0 | 0 |
| NGC 450 | | | | 0.95-1.0 | < 0.1 | < 0.1 |
| NGC 1739 | 0.65 | 0.3-0.4 | 0.24-0.3 | 0.55 | 0.2-0.26 | 0.16 |
| NGC 4568 | 0.5-0.9 | 1.1 | 0.69 | | | |
| NGC 3314 | 0.16 | 1.6 | 1.24 | 0.19 | 1.11 | 1.60 |
| " | 0.34 | 1.64 | 0.82 | 0.28 | 0.77 | 0.59 |
| " | 0.42 | 1.11 | 0.82 | 0.39 | 1.75 | 0.63 |

Figure 3. Extinction measures in arm and interarm regions. Values of $A_B$ and $A_I$ are plotted as blue and red diamonds, respectively.